%% file: main.tex
\documentclass[runningheads]{template_resources/llncs}
\usepackage{graphicx}

\input{math_commands}

\begin{document}

%
\title{Sequential Nature of Recommender Systems Disrupts the Evaluation Process}
%
%
\author{Ali Shirali\inst{1}\orcidID{0000-0003-3750-0159}}
\authorrunning{A. Shirali}
%
\institute{University of California, Berkeley, CA, USA \\
\email{shirali\_ali@berkeley.edu}}
\maketitle              

\input{sections/abstract}

\input{sections/intro}

\input{sections/eval_and_adapt}
\input{sections/knn_recsys}
\input{sections/seq_attack}
\input{sections/experiments}
\input{sections/discussion}

%
%
%

\bibliographystyle{template_resources/splncs04}
\bibliography{refs}

\end{document}

%% file: math_commands.tex
\usepackage{amsmath,amsfonts,bm}

\def\eqref#1{equation~\ref{#1}}

\def\1{\bm{1}}

\def\vone{{\bm{1}}}

\def\vb{{\bm{b}}}

\def\vi{{\bm{i}}}

\def\vu{{\bm{u}}}
\def\vv{{\bm{v}}}

\def\vx{{\bm{x}}}
\def\vy{{\bm{y}}}

\def\mX{{\bm{X}}}

\DeclareMathAlphabet{\mathsfit}{\encodingdefault}{\sfdefault}{m}{sl}
\SetMathAlphabet{\mathsfit}{bold}{\encodingdefault}{\sfdefault}{bx}{n}

\def\gA{{\mathcal{A}}}

\def\gC{{\mathcal{C}}}
\def\gD{{\mathcal{D}}}

\def\gI{{\mathcal{I}}}
\def\gJ{{\mathcal{J}}}

\def\gN{{\mathcal{N}}}

\def\gU{{\mathcal{U}}}

\def\gW{{\mathcal{W}}}

\def\gY{{\mathcal{Y}}}

\newcommand{\E}{\mathbb{E}}

\newcommand{\R}{\mathbb{R}}



%% file: sections/abstract.tex
\begin{abstract}
Datasets are often generated in a sequential manner, where the previous samples and intermediate decisions or interventions affect subsequent samples. This is especially prominent in cases where there are significant human-AI interactions, such as in recommender systems. To characterize the importance of this relationship across samples, we propose to use adversarial attacks on popular evaluation processes. We present sequence-aware boosting attacks and provide a lower bound on the amount of extra information that can be exploited from a confidential test set solely based on the order of the observed data. We use real and synthetic data to test our methods and show that the evaluation process on the MovieLense-100k dataset can be affected by $\sim1\%$ which is important when considering the close competition. Codes are publicly available. \footnote{\url{https://github.com/alishiraliGit/augmented-boosting-attack}}

\keywords{Sequential Decision Making  \and Sequential Recommender System \and Evaluation Mechanisms \and Missing-Not-At-Random}
\end{abstract}

%% file: sections/intro.tex
\section{Introduction}
Datasets are frequently generated in a sequential manner, where there is substantial shared information between samples. Particularly, in settings where there are significant human-AI interactions, decisions made by algorithms often take place sequentially based on observed feedback from previous decisions. Through this process, algorithms can learn more about individuals they are interacting with and indirectly also learn about other similar individuals. For example, consider a patient whose response to a current drug provides information about the specific disease the patient had and what may provide better treatment for similar individuals in the future. Similarly, how a user rates a movie informs recommender systems about that individual's interest and other users in the same demographic group. 

Despite the ubiquity of sequentially-generated datasets, there remains room for understanding how significant of a challenge this sequential nature presents to the pipeline of training and evaluating a model. In this work, we specifically focus on the evaluation process and show how it might be distorted by samples that are not independently generated. 

In a general prediction task, we train and evaluate a predictor by setting aside a part of the data as the \textit{holdout} or \textit{test} set, and train the predictor on the rest of the data (\textit{training} set). We then evaluate the trained model using \textit{empirical risk} on the test set as a proxy for the true risk. When samples in the test set are independently and identically distributed (i.i.d), this empirical risk will be close to the true risk with high probability. But this no longer holds if the samples are not i.i.d, which is the case in sequentially-generated datasets. We focus on the gap between this empirical risk on the test set and the true risk to show the extent our evaluation might be disrupted.

To characterize the role of sequential information in the evaluation process, we propose to use \textit{adversarial} attacks on the test set with and without the knowledge of the data generation order. Generally, an adversarial attacker tries to reveal the test set by sending many queries to the evaluation system and adaptively designing new predictions. Such an attacker is blind to the training set and cannot do better than random in terms of the true risk. What if, however, the attacker was given sequential information?

We explore this question using recommender systems (RS), where sequentially-generated datasets are commonplace. In fact, the major benchmark datasets of recommendation were collected when another RS, \textit{logging} RS, was frequently interacting with users. As the logging RS tried to offer favorable items to users while learning their preference, the collected samples are neither complete random drawn from user-item space nor independent of each other. First of all, observed ratings are often positively biased. In other words, users are exposed to recommendations that are more likely to be in their favor compared to a set of random recommendations. This effect is studied under the Missing-Not-At-Random (MNAR) problem~\cite{little2019statistical,pradel2012ranking}. Second, the order of observation informs us beyond just being likely to be positive. Change or consistency in the category of recommended items over time can be a sign of dislike or like. For example, if a horror movie is recommended to a user after a comedy movie, we may infer that the user was not interested in comedies. Here, the observation order matters because the opposite conclusion could be drawn if the comedy movie was recommended after the horror. 

To exploit the information hidden in the order of observed data, we need full knowledge of the logging RS algorithm. However, this knowledge is unlikely to be openly available. We, therefore, propose a simple $k$-NN RS to approximate the logging RS. The $k$-NN RS is simple compared to the state-of-the-art algorithms, but we will show it effectively approximates a real logging RS.

Although we have focused on adversarial attacks trying to reveal the test set, information leakage might naturally happen when an adaptive algorithm tries to improve its predictions after observing the performance of previous predictions. This adaptation harms the predictor's generalization ability and disrupts the relative performance evaluation of multiple algorithms (for example, in a competition). As a response, a natural evaluation mechanism, the Ladder~\cite{ladder}, is suggested which for any new submission reports empirical risk on the test set only if the risk is improved. The ladder blocks too much information leakage due to multiple submissions. We will study the importance of sequential information under conventional and ladder evaluation mechanisms. 

In the following, after reviewing related works and introducing the notation, we first formalize the evaluation process and adversarial attacks in Section~\ref{sec:eval_and_adapt}. We then study recommender systems as sequential decision-makers and introduce $k$-NN RS in Section~\ref{sec:knn_recsys}. We propose two sequence-aware boosting attacks in Section~\ref{sec:seq_attack}. The experiments and results are then discussed in Section~\ref{sec:exp}.

\subsection{Related Works}
The human-RS interaction gets more complicated when we consider the sequential nature of human decision-making. Human decisions might be directed towards a goal. This leads to \textit{complementary} preference over items. For example, a user searching for an airplane ticket will probably look for a hotel as well. Various methods are proposed to capture complementary preference over items, including but not limited to Markov chains~\cite{shani2005mdp,garcin2013personalized,he2016fusing}, sequence-aware factorization machines~\cite{rendle2010factorizing}, and Recurrent Neural Networks (RNN)~\cite{hidasi2015session,wu2017recurrent}. We distinguish the complexity of sequential decision-making of humans from the complexity of sequential decisions of RS. In the given example, we might study flight and hotel recommenders separately. Note that the recommendation process for the selected topic is still sequential because the recommender is learning the user's preference. 

The logging RS attempts to recommend favorable items, resulting in positively biased ratings. This problem is generally studied under Missed-Not-At-Random (MNAR) problem~\cite{pradel2012ranking,little2019statistical}. Training a model on MNAR data usually yields overly optimistic predictions over unseen ratings. A direct way to address MNAR is to treat the recommendation problem as missing-data imputation of the rating matrix based on the joint likelihood of the missing data model and the rating model~\cite{hernandez2014probabilistic}. This leads to sophisticated methods. Alternatively, Inverse Propensity Scoring (IPS) has been shown to effectively debias training and evaluation on MNAR data~\cite{schnabel2016recommendations,wang2019doubly}. In IPS, each term of the empirical risk corresponding to an observed rating will be normalized by the inverse of the probability they would have been observed (aka propensity). Existing works typically estimate propensities as outputs of a simple predictor such as logistic regression or naive Bayes~\cite{schnabel2016recommendations} or more recently reconstruct it under low nuclear norm assumption~\cite{ma2019missing}. In none of these works, sequential information is exploited. However, we will show that sequential information is effective and important in the rating prediction task and consequently can be used for better propensity estimation.

\subsection{Notation and Definitions}
We use lowercase, bold-faced lowercase, and bold-faced uppercase letters to represent scalars, vectors, and matrices.

Focusing on the binary classification task, we represent $i^{th}$ sample by its \textit{feature} vector $\vx^{(i)} \in \R^d$ and \textit{label} $y^{(i)} \in \{0, 1\}$. For any set $S$ of samples, we can put together feature vectors as rows of a matrix and labels as elements of a vector to obtain $\mX_S \in \R^{|S| \times d}$ and $\vy_S \in \{0, 1\}^{|S|}$, respectively. We use $\gD$ to refer to the training data $(\mX_{train}, \vy_{train})$. 

Unless otherwise stated, we use zero-one \textit{loss} function as the performance indicator of a prediction: $loss(\hat{y}, y) = 1_{\hat{y} = y}$. The empirical risk of a classifier $f: \R^d \rightarrow \{0, 1\}$ over samples $S$ is defined as $R_S[f] = \frac{1}{|S|} \sum_{i \in S} loss(f(\vx^{(i)}), y^{(i)})$. Assuming features and labels are drawn from a joint distribution $P$, the (true) risk of a classifier is defined as $R[f] = \E_{\vx, y \sim P}[loss(f(\vx), y)]$. We sometimes explicitly refer to the empirical and true risk of predicted labels $\hat{\vy}$ as $R_S[\hat{\vy}]$ and $R[\hat{\vy}]$, respectively.

%% file: sections/eval_and_adapt.tex
\section{Evaluation Systems and Adaptation}
\label{sec:eval_and_adapt}

\begin{figure}[t]
    \centering
    \includegraphics[width=0.8\linewidth]{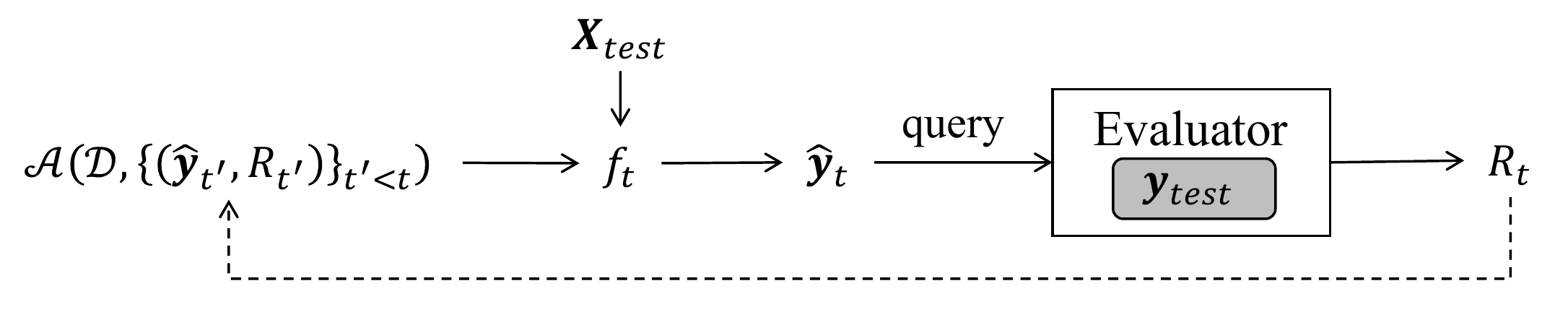}
    \caption{Training and evaluation process of an adaptive classifier.}
    \label{fig:eval_process}
\end{figure}

Generally, we can define three interacting components in the evaluation process of a classification task: data, evaluator, and algorithm as shown in Figure \ref{fig:eval_process}. In the holdout method, data is divided into training ($\gD$) and test (holdout) sets, while labels corresponding to the test set ($\vy_{test}$) are secured in the evaluator and hidden to the algorithm. At each time step $t$, the evaluator compares the input predictions $\hat{\vy}_t$ to $\vy_{test}$ and reports a performance indicator $R_t$. For example, the \textit{Kaggle} mechanism reports the empirical risk $R_{test}[\hat{\vy}_t]$ with $10^{-5}$ precision. The evaluator can have a state and might not necessarily report the empirical risk. 

A non-adaptive algorithm trains a classifier $f_t$ on $\gD$ regardless of the previously reported performances. If samples in the test set are generated i.i.d, applying Hoeffding's bound and union bound implies the empirical risk on the holdout set is close to the true risk with high probability:
\begin{equation}
\label{eq:bound_non_adaptive}
    P(\exists t \le T: |R_{test}[f_t] - R[f_t]| > \epsilon) \le 2T \exp{(-2\epsilon^2 n_{test})}
\end{equation}
where $n_{test}$ is the number of samples in the test set. However, even when samples are i.i.d., an adaptive algorithm might use the performance on the previous predictions in training phase to design a new predictor:
\begin{equation}
    f_t = \gA(\gD, \{(\hat{\vy}_{t'}, R_{t'})\}_{t' < t})
    .
\end{equation}
In this case, $f_t$ is a function of all samples in the test set, so $loss(\hat{y}_t^{(i)}, y^{(i)})$ is not independent of other losses and Hoeffding's bound is not applicable. Consequently, the bound from Equation \ref{eq:bound_non_adaptive} is no longer valid and empirical risk on the test set might be very far from the true risk~\cite{ladder}. 

\subsection{Ladder Mechanism}
The Ladder mechanism~\cite{ladder} keeps the best (smallest) empirical risk so far achieved as its state. As long as new predictions do not improve the best seen risk, it refuses to report their empirical risks. Formally, the Ladder's state at time $t$ is $R_t^{best}$. Let $[.]_\eta$ operator rounds the input to the nearest integer multiple of $\eta$. For any new classifier $f_t$, Ladder returns $R_t^{best} = [R_{test}[f_t]]_\eta$ if $R_{test}[f_t] < R_{t-1}^{best} - \eta$ and $R_t^{best} = R_{t-1}^{best}$ otherwise. Even when classifiers are trained adaptively, at any time $t$, $R_t^{best}$ will remain close to the best empirical risk observed so far (Theorem 3.1 of~\cite{ladder}) which means $R_t^{best}$ is unlikely to diverge significantly from $\min_{t' \le t} R[f_{t'}]$.

\subsection{Adversarial Attacks}
An attacker tries to reveal $\vy_{test}$ by many queries to the evaluator without actually learning anything about the patterns of data. A well-known randomized attack is \textit{boosting attack} where at each time step $t$ the attacker
\begin{enumerate}
    \item Query evaluator with a random vector $\vv_t \sim unif(\{0, 1\}^{n_{test}})$.
    \item If $R_t$ returned by the evaluator is better than random and better than the best risk observed so far, add $\vv_t$ to the set of informative vectors $\gJ$.
\end{enumerate}
The final prediction of the attacker is the elementwise majority vote of all informative vectors: $\vv^* = majority(\{\vv_t \in \gJ\})$.

\begin{figure}[t]
    \centering
    \includegraphics[width=0.8\linewidth]{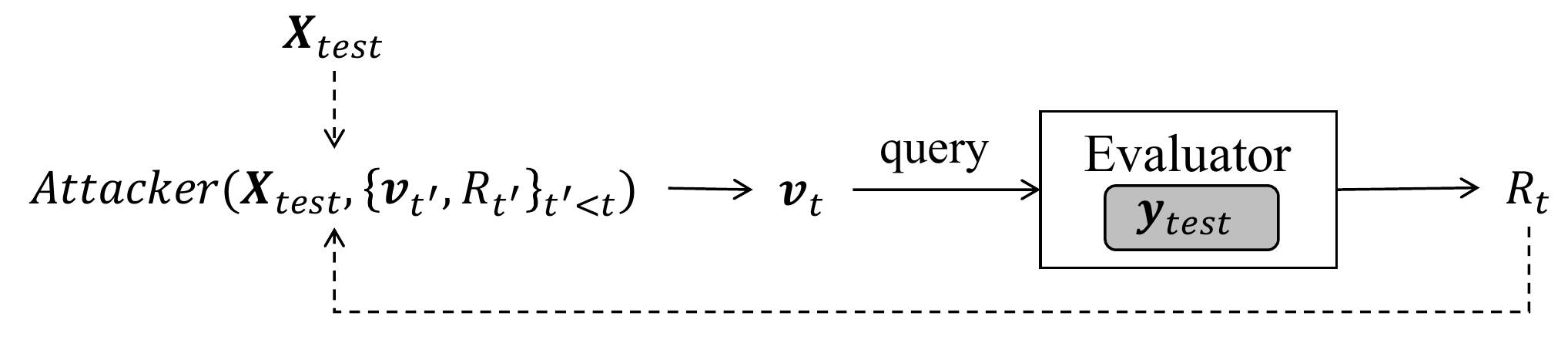}
    \caption{A general attacker might use $\mX_{test}$ and previous risks and queries for the new guess.}
    \label{fig:general_attacker}
\end{figure}

Queries are generally random, but they might also depend on previous queries (Figure~\ref{fig:general_attacker}). For example, queries of the boosting attack are generated uniformly at random, regardless of the previous queries. However, in the random window boosting attack we propose in Section~\ref{subsec:wboost}, the distribution of a new query depends on the queries and risks observed so far.

Attackers do not use the training data, as they would otherwise be real learners, but they might use $\mX_{test}$. For example, the boosting attack is unaware of $\mX_{test}$, but the sequence-aware attackers we propose in Section~\ref{sec:seq_attack} use $\mX_{test}$ to elicit information about observation order.

%% file: sections/knn_recsys.tex
\section{Recommender Systems as Sequential Decision Makers}
\label{sec:knn_recsys}

\begin{figure}[t]
    \centering
    \includegraphics[width=0.8\linewidth]{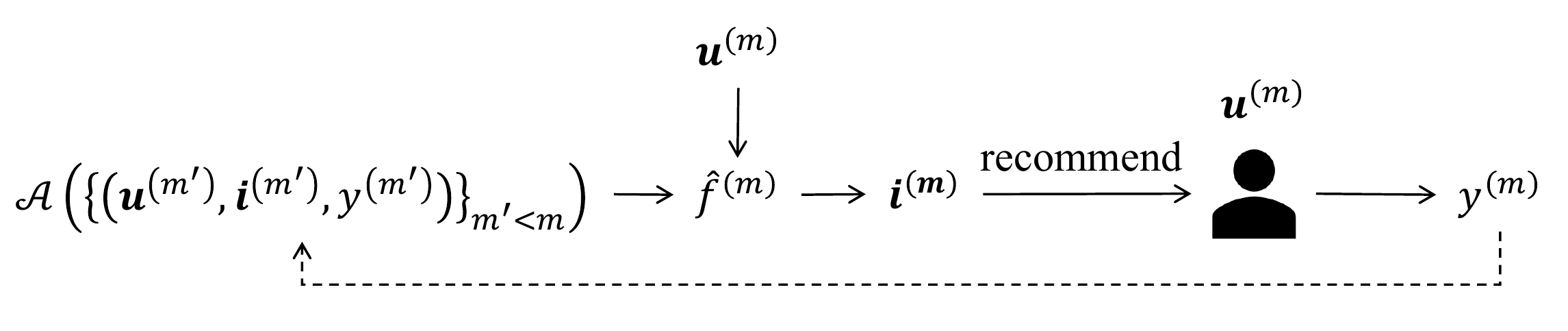}
    \caption{The process of recommendation for a general RS.}
    \label{fig:general_recsys}
\end{figure}

Generally, an RS consists of a set of users ($\gU$) and items ($\gI$). We assume a user (an item) can be represented by a vector $\vu \in \gU$ ($\vi \in \gI$). During the recommendation process (Figure~\ref{fig:general_recsys}), at each time $m$, a random user $\vu^{(m)}$ asks for a recommendation. The RS has a trained classifier $\hat{f}^{(m)}: \gU \times \gI \rightarrow \{1, 0\}$, which predicts whether a user will like an item or not. So, the RS suggests item $\vi^{(m)}$ randomly selected from all items that satisfy $\hat{f}^{(m)}(\vu^{(m)}, \vi) = 1$ and observes the user's feedback $y^{(m)} = f(\vu^{(m)}, \vi^{(m)}) \in \{1, 0\}$. Here, we assumed feedback is completely determined by $(\vu, \vi)$ and is time-invariant. This assumption implicitly means user preferences do not change over time and the vector representations of users and items are enough to determine the feedback. In practice, these assumptions are only valid with approximation. Based on the observed feedback, RS updates its classifier $\hat{f}^{(m+1)} = \gA(\{\vu^{(m')}, \vi^{(m')}, y^{(m')}\}_{m' < m})$. So, how a user responded to the previous recommendations impacts future recommendations. 

\subsection{$k$-NN Recommender System}
The $k$-NN RS is a simple RS we use to approximate a real logging RS. For any pair of user-item $(\vu, \vi)$, the $k$-NN RS uses a $k$-NN classifier to assign a label to $(\vu, \vi)$ based on the majority of feedback already observed from $k$ nearest user-items. 

Formally, let $\gN_k^{(m)}(\vu, \vi)$ be the set of $k$ time stamps $\{m'\}$ corresponding to $\{(\vu^{(m')}, \vi^{(m')})\}$, which have the shortest distance to $(\vu, \vi)$ among all the ratings already observed till time $m$ ($m' < m$). We measure distance by $\| \vu - \vu^{(m')} \|^2_2 + \| \vi - \vi^{(m')} \|^2_2$. Then
\begin{equation}
\label{eq:f_hat}
    \hat{f}^{(m)}(\vu, \vi) = majority(\{y^{(m')} \mid m' \in \gN_k^{(m)}(\vu, \vi)\})
    .
\end{equation}

To control the level of divergence from being i.i.d., we introduce a new parameter named $exploration \in [0, 1]$. For any new user $\vu^{(m)}$, we define $\gI^{(m)}_+ = \{\vi \in \gI \mid \hat{f}^{(m)}(\vu^{(m)}, \vi) = 1\}$ and $\gI^{(m)}_- = \{\vi \in \gI \mid \hat{f}^{(m)}(\vu^{(m)}, \vi) = 0\}$, where we assume $|\gI|$ is finite. Then RS suggests an item $\vi$ from $\gI = \gI^{(m)}_+ \cup \gI^{(m)}_-$ such that 
\begin{equation}
\label{eq:I_plus_minus}
    \frac{P(i \in \gI^{(m)}_-)}{P(i \in \gI^{(m)}_+)} = exploration \times \frac{|\gI^{(m)}_-|}{|\gI^{(m)}_+|}
\end{equation}
So, $exploration=1$ corresponds to having no preference over $\gI^{(m)}_-$ and $\gI^{(m)}_+$, and recommendations are perfectly random. In contrast, $exploration=0$ corresponds to the case where only items predicted to receive positive feedback will be recommended. 

Finally, putting in the context of a classification problem, $y^{(m)}$ is the label for feature vector $\vx^{(m)} = (\vu^{(m)}, \vi^{(m)}, m)$. We explicitly include time $m$ in the feature vector to investigate the value of this extra information later. 

%% file: sections/seq_attack.tex
\section{Sequence-Aware Adversarial Attacks}
\label{sec:seq_attack}
This section proposes two augmented boosting attacks specifically targeted for sequential data. The boosting attack queries the evaluator with completely random vectors from $\{0, 1\}^{n_{test}}$ because it doesn't have any prior on the labels of the test set. However, if samples are generated sequentially, and the order of their generation is accessible, we can update our prior and design better queries.  

We incorporate order information in two different ways, resulting in the following algorithms. The first algorithm (Section~\ref{subsec:wboost}) is a model-free algorithm based on the intuition that initial recommendations to a user are less likely than the following recommendations to receive positive feedback. So, over time, there is a distribution shift that the proposed random window boosting attack (\textit{WBoost}) tries to learn. In contrast to the boosting attack, the next query in this method depends on the previous queries and responses.

The second proposed algorithm (Section~\ref{subsec:post_attack}) is a model-based algorithm that considers $k$-NN RS as an approximation of the logging RS. Based on the order of the observed data, it calculates the posterior probability over the test set's labels. Compared to the boosting attack, this method samples queries according to this posterior probability rather than uniform distribution, which significantly increases the chance of guessing the correct labels.

\subsection{Random Window Boosting Attack (WBoost)}
\label{subsec:wboost}
In this method, we try to compensate for the distribution shift of labels over time. The discussed boosting attack is blind to this shift because it simply samples query from $\vv \sim unif(\{0, 1\}^{n_{test}})$ without considering that elements of $\vy_{test}$ are recommended at different times. 

\begin{figure}[t]
    \centering
    \includegraphics[width=0.95\linewidth]{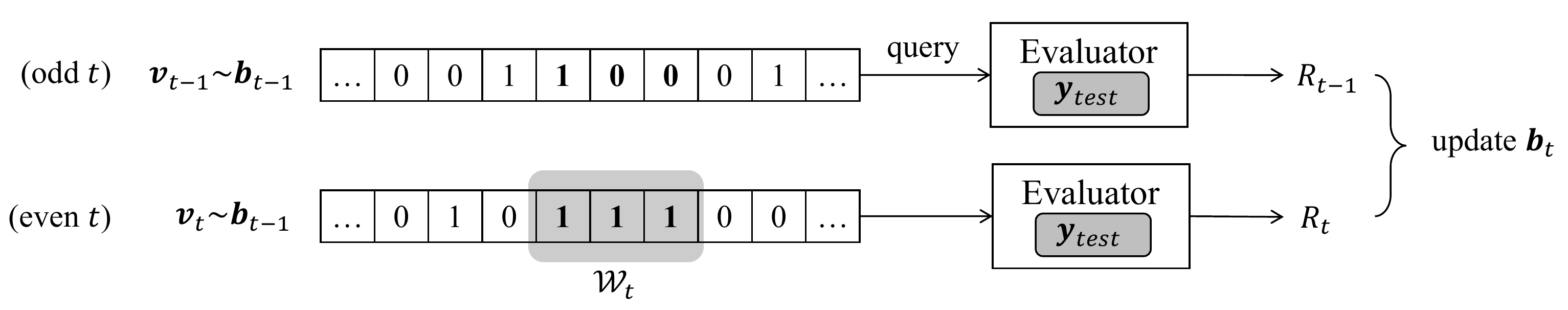}
    \caption{An example from the WBoost updating procedure}
    \label{fig:wboost}
\end{figure}

Without loss of generality, we assume labels in $\vy_{test}$ are ordered in time (the order that items are recommended is available in $\mX_{test}$ so we can rearrange elements of $\vy_{test}$ chronologically). We define the state of the attacker at time $t$ with $\vb_t \in [0, 1]^{n_{test}}$ where $b_t^{(m)}$ ($m^{th}$ element of $\vb_t$) is the attacker's estimation from $P(y_{test}^{(m)} = 1)$. The algorithm starts with $\vb_1 = \frac{1}{2}\vone$ (a vector of all $0.5$) at time $t = 1$. At odd time $t$, it generates $\vv_t$ according to $\vb_t$. At even time $t$, it selects a random window with length $w$ from $\vv_{t-1}$, assigns the value of all elements under the window to the their minority value, and draws the rest of the elements from $\vb_{t-1}$. Then by observing $R_t$ from the evaluator, the attacker updates $\vb_t$. Figure~\ref{fig:wboost} shows an example of this process.

Here is the updating rule: let $\gW_t$ be elements under the window at even time $t$. As elements outside of the window in both $\vv_{t-1}$ and $\vv_t$ are selected according to $\vb_{t-1}$, in expectation, they don't have any effect on $R[\vv_t] - R[\vv_{t-1}]$. Let's assume the actual probability of $P(y_{test}^{(m)} = 1)$ is $B^*$ for all $m \in \gW_t$, then:
\begin{align}
\label{eq:b_star}
    \E[R[\vv_t] - R[\vv_{t-1}]] &=
    \frac{1}{n_{test}} \sum_{m \in \gW_t} P(y_{test}^{(m)} = 1) (1 - v_t^{(m)}) + P(y_{test}^{(m)} = 0) v_t^{(m)} \nonumber \\
    &- \frac{1}{n_{test}} \sum_{m \in \gW_t}  P(y_{test}^{(m)} = 1) (1 - v_{t-1}^{(m)}) + P(y_{test}^{(m)} = 0) v_{t-1}^{(m)}
    \nonumber \\
    &= \frac{1 - 2 B^*}{n_{test}} \sum_{m \in \gW_t} v_t^{(m)} - v_{t-1}^{(m)}
\end{align}
Where expectation is taken w.r.t elements outside of the $\gW_t$. Using $R_t$ and $R_{t-1}$ returned by the evaluator as a proxy for $\E[R[\vv_t]]$ and $\E[R[\vv_{t-1}]]$ we can estimate $B^*$ and update $\vb$:
\begin{equation}
    b_t^{(m)} = \begin{cases}
        (1 - \alpha) b_{t-1}^{(m)} + \alpha \frac{1}{2} \Big(1 - n_{test} \frac{R_t - R_{t-1}}{\sum_{m \in \gW_t} v_t^{(m)} - v_{t-1}^{(m)}} \Big) 
        & m \in \gW_t \\
        b_{t-1}^{(m)} & o.w.
    \end{cases}
\end{equation}
where $\alpha \in (0, 1]$ controls the speed of the convergence.

\subsection{$k$-NN Posterior Boosting Attack (PostBoost)}
\label{subsec:post_attack}
In this method, we approximate the logging RS with a $k$-NN RS and calculate a posterior probability over the unseen test set's labels based on available test set's features: $P(\vy_{test} = \vv \mid \mX_{test})$. Again, we presume elements of $\vy_{test}$ and $\vv$ are ordered chronologically (o.w. we rearrange them). 

\begin{figure}[t]
    \centering
    \includegraphics[width=0.85\linewidth]{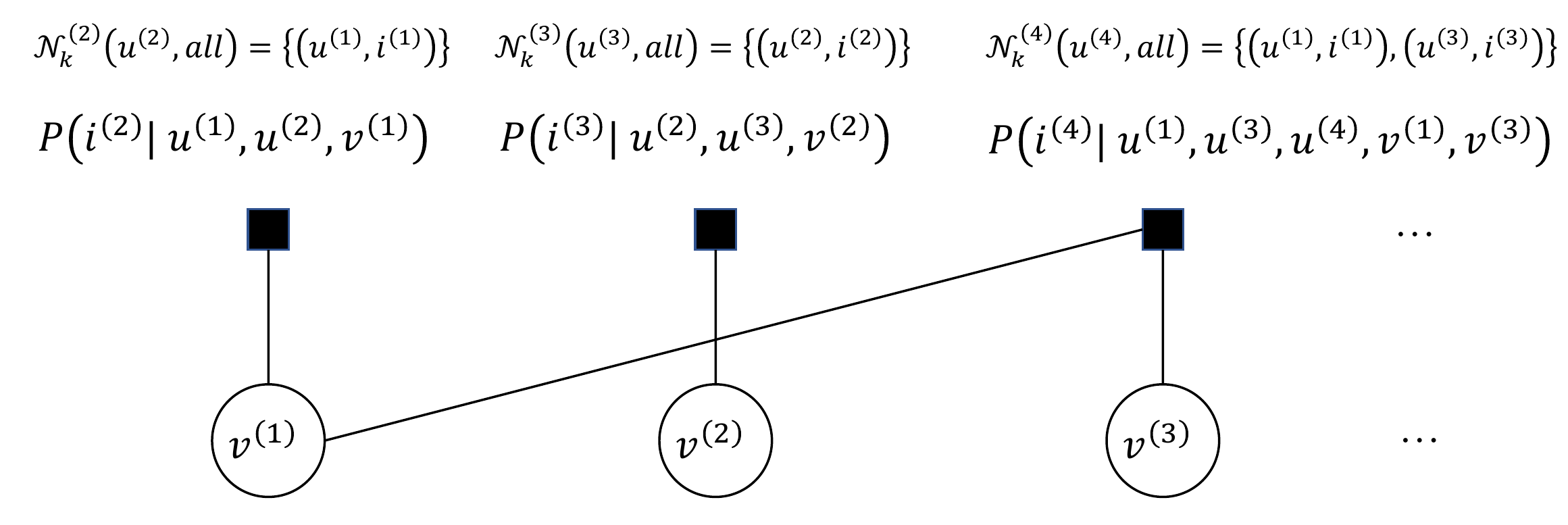}
    \caption{An example  factor graph utilized in calculating posterior probability over test set's labels.}
    \label{fig:factor_graph}
\end{figure}

As defined in Section~\ref{sec:knn_recsys}, let $\gN_k^{(m)}(\vu, \vi)$ be the set of $k$ nearest neighbours of $(\vu, \vi)$ from $\{(\vu^{(m')}, \vi^{(m')})\}_{m' < m}$. We define $\gN_k^{(m)}(\vu, all) = \bigcup_{\vi \in \gI} \gN_k^{(m)}(\vu, \vi)$ as the set of all previous recommendations which are important in determining the next recommendation to user $\vu$. 
At any time $m$, for a given observed ratings $\{y^{(m')}=v^{(m')}\}_{m' < m}$, we can find the RS's classifier ($\hat{f}^{(m)}$) from Equation~\ref{eq:f_hat}. Then we can calculate the likelihood that $\vi^{(m)}$ is recommended next according to Equation~\ref{eq:I_plus_minus}:
\begin{align}
    P_m &= P\Big(\vi^{(m)} \mid \vu^{(m)}, \big\{(\vu^{(m')}, \vi^{(m')}, y^{(m')}=v^{(m')}) \mid m' \in \gN_k^{(m)}(\vu^{(m)}, all)\big\} \Big) \nonumber \\
    &= \begin{cases}
        \frac{exploration}{exploration \times |\gI_-^{(m)}| + |\gI_+^{(m)}|} & \hat{f}^{(m)}(\vu^{(m)}, \vi^{(m)}) = 0 \\ 
        \frac{1}{exploration \times |\gI_-^{(m)}| + |\gI_+^{(m)}|} & \hat{f}^{(m)}(\vu^{(m)}, \vi^{(m)}) = 1
    \end{cases} 
    .
\end{align}
So, we can calculate the likelihood
\begin{equation}
\label{eq:map_likelihood}
    P(\{\vi^{(m)}\}_{m=1}^{n_{test}} \mid \{\vu^{(m)}\}_m, \vy=\vv) 
    = \prod_{m=1}^{n_{test}} P_m
    .
\end{equation}
for any $\vv$. Without loss of generality, we assume data is balanced and the prior $P(v^{(m)}) = 0.5$. Now, we can find the posterior $P(\vy=\vv \mid \{(u^{(m)}, i^{(m)})\}_{m=1}^{n_{test}}) = P(\vy=\vv \mid \mX_{test})$ from the likelihood of Equation~$\ref{eq:map_likelihood}$. Here, we utilize a factor graphical model, with elements of $\vv$ as variable vertices and $P_m$s as factor vertices. The marginal joint distribution of variable vertices, will be our desired $P(\vy=\vv \mid \mX_{test})$. An example of this graph is depicted in Figure~\ref{fig:factor_graph}.

\subsubsection{Approximate Posterior Calculation}
When $n_{test}$ is large, finding the posterior probability over the test set's labels is not computationally feasible. So, we propose to cluster ratings into equal size groups and find the posterior probability for each cluster separately. In order to do a sequence-aware clustering, we start by $\gC_0 = [n_{test}]$ and iteratively find clusters. For a desired cluster size $z$, at each iteration $t \ge 1$, we
\begin{enumerate}
    \item Draw a random $m$ from $\gC_{t-1}$.
    \item Select $z - 1$ members from $\gN_k^{(m)}(\vu^{(m)}, \vi^{(m)}) \cap \gC_{t-1}$ randomly.
    \item Form a new cluster with $m$ and the selected $z-1$ neighbours.
    \item Remove the new cluster's members from $\gC_{t-1}$ and obtain $\gC_t$.
\end{enumerate}

%% file: sections/experiments.tex
\section{Experiments}
\label{sec:exp}
To investigate the importance of sequential information in real scenarios, we use MovieLense-100k (ML-100k)~\cite{ml-100k}, a well-known benchmark dataset for rating prediction. What makes ML-100k interesting for us is the availability of timestamps for samples. Besides ML-100k, we also use synthetic data to study the effectiveness of the proposed methods in a controlled environment with the desired level of divergence from randomness. 

The common practice in evaluating recommender systems is to select the test set randomly from all observations, regardless of their order. The ML-100k has prespecified test sets that follow a similar practice. We also use a similar method in setting aside the test set from the synthetic data. Whether data is obtained from real interactions or from simulation, it might be biased towards positive ratings. So, we downsample the initial test set to have a balanced test set. The balanced test set reassures the effects we will observe are due to the sequential nature of the data. 

In all experiments, attackers do not have access to any training data and labels. So, the true risk of an attacker cannot be better than the chance level, which is $0.5$ in our balanced design. Therefore, the difference of the attacker's average loss on the test set with $0.5$ reflects the amount of information leaked about the test set. As the boosting attack is unaware of the sequential nature of the data, the gap between a sequence-aware attacker and the boosting attack shows the extent that sequential information can disrupt the evaluation. The level of disruption might vary based on the evaluation mechanism. Notably, we compare two natural mechanisms, Kaggle and Ladder, in this regard.

\paragraph{Synthetic data simulation.}
In simulating ratings, we assume that the user's feedback function ($f$) follows a similar structure as Equation~\ref{eq:f_hat} and specify it with ground truth centers and labels: $\gC_* = \{(\vu_*^{(l)}, \vi_*^{(l)})\}_l$ and $\gY_* = \{y_*^{(l)}\}_l$. For any pair of user and item $(\vu, \vi)$, let $\gN_k^*(\vu, \vi)$ be the set of $k$ closest $(\vu_*^{(l)}, \vi_*^{(l)}) \in \gC_*$ to $(\vu, \vi)$. Then the label of $(\vu, \vi)$ will be
\begin{equation}
    f^*(\vu, \vi) = majority(\{y^{(l)}_* \mid l \in \gN_k^*(\vu, \vi)\})
    .
\end{equation}
Given the feedback function $f^*$, we run another $k$-NN RS as the logging RS to obtain samples: at each time $m$, a random user $\vu^{(m)}$ queries the logging RS and receives $\vi^{(m)}$. The user's response to $\vi^{(m)}$ will be determined by $f^*(\vu^{(m)}, \vi^{(m)})$.

\subsection{Evaluation on Synthetic Data}
\begin{figure}[t]
    \centering
    \includegraphics[width=0.8\linewidth]{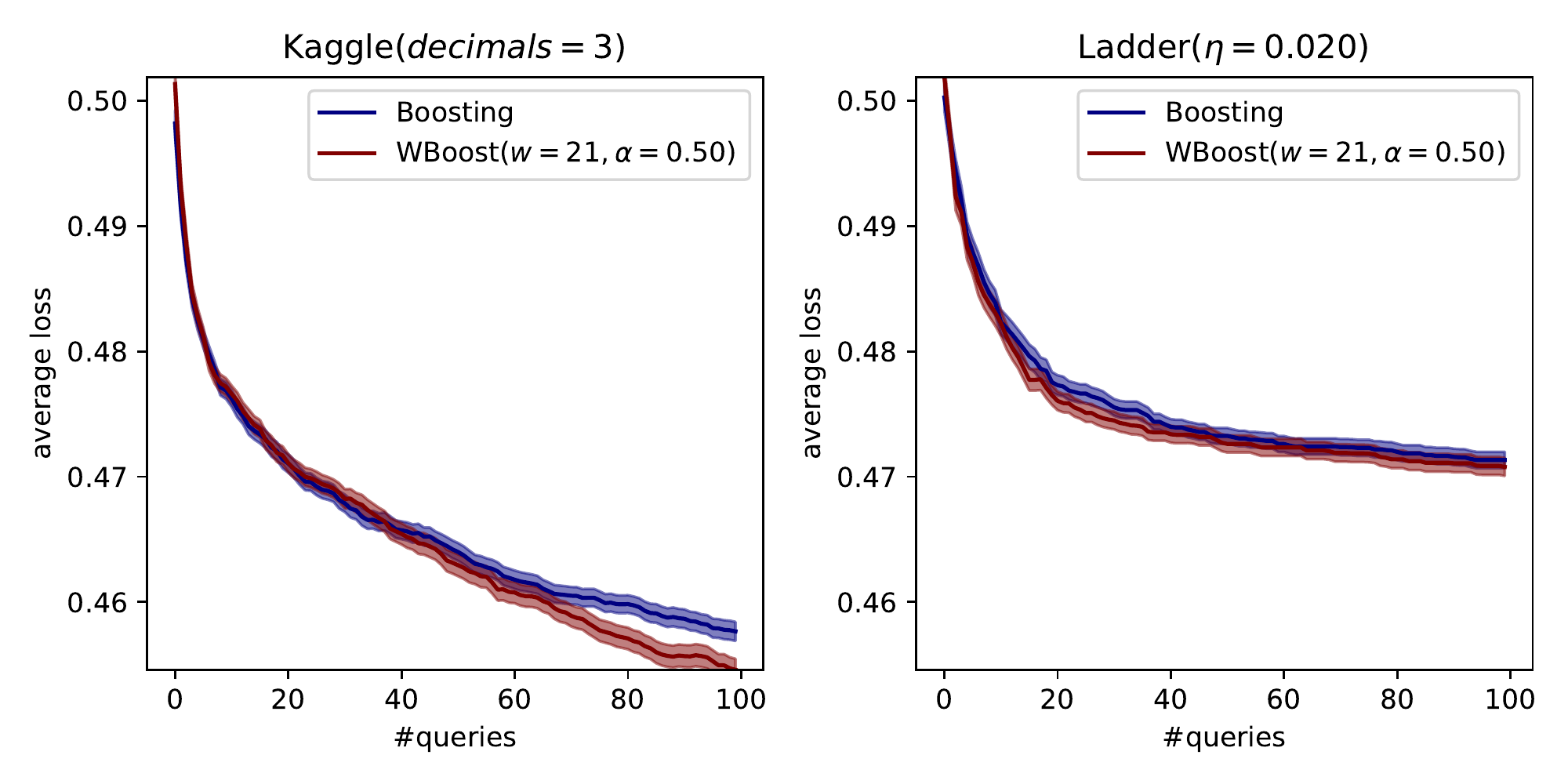}
    \caption{Empirical risk on the test set of synthetic data with $1000$ samples while WBoost sends more queries to evaluators.}
    \label{fig:wboost_on_synthetic}
\end{figure}

We simulate synthetic data consisting of $1000$ balanced samples obtained by a $1$-NN logging RS to investigate how well the random window boosting attack (WBoost) can exploit the sequential information. Figure~\ref{fig:wboost_on_synthetic} shows the empirical risk on the test set while the attackers send more and more queries to the evaluator. Here we have repeated data simulation and attacks and plotted average results. Note that the value on the y-axis (empirical risk or equivalently average loss) is not accessible to attackers; what attackers observe depends on the evaluation mechanism (e.g., Kaggle reports the empirical risk with a limited precision). One can see that WBoost gradually learns the distribution shift over time when the evaluator is Kaggle; however, the Ladder mechanism effectively blocks it.

\begin{figure}[t]
    \centering
    \includegraphics[width=0.8\linewidth]{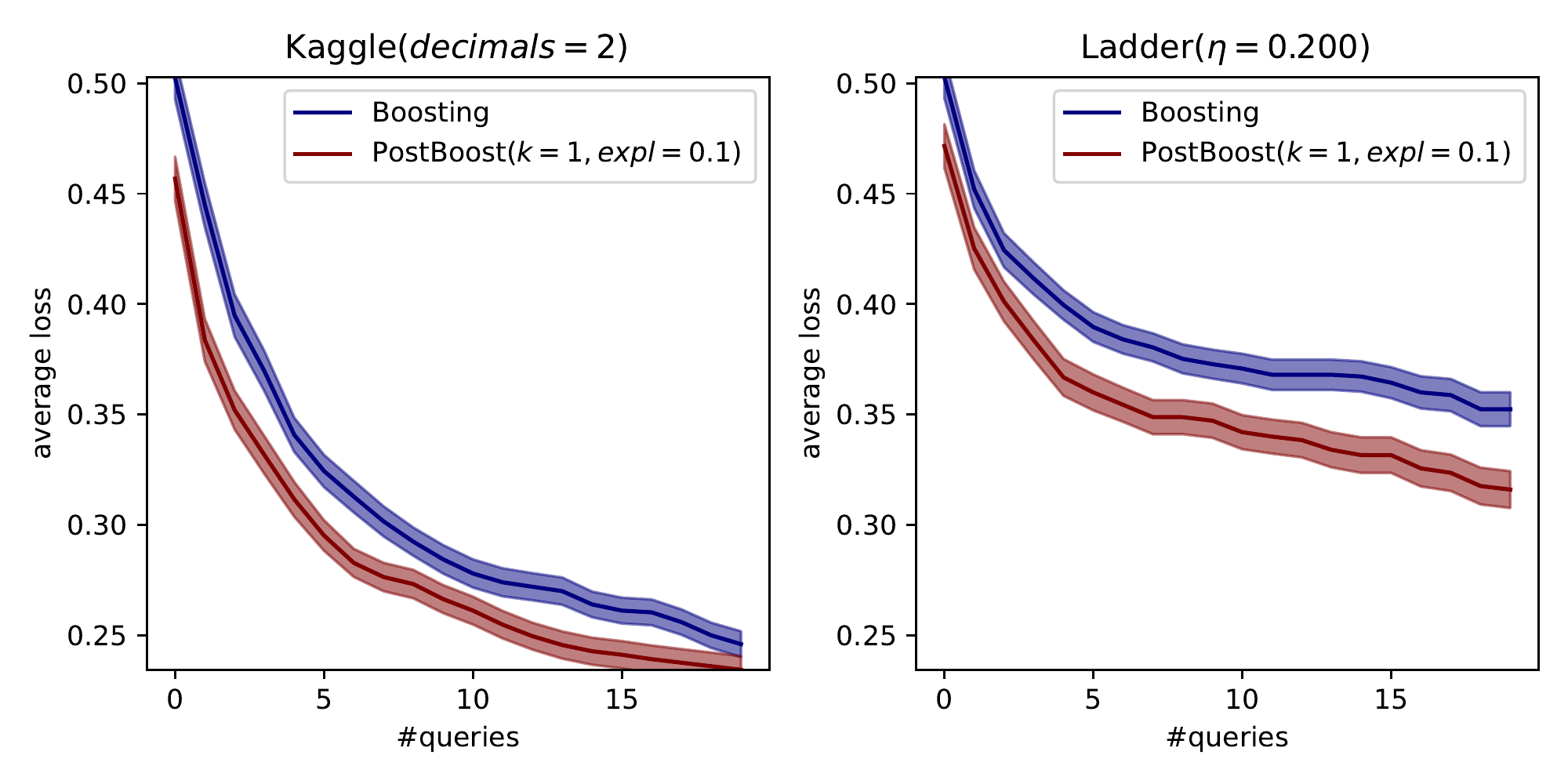}
    \caption{Empirical risk on the test set of synthetic data with $10$ samples while $k$-NN posterior boosting attacker (PostBoost) sends more queries to evaluators.}
    \label{fig:post_boost_synthetic}
\end{figure}

We simulate a very smaller data consisting of only $10$ balanced samples to observe the full capability of $k$-NN posterior boosting attack (PostBoost). The computational complexity of PostBoost limits the number of samples it can infer about. Although the approximate PostBoost explained in Section~\ref{subsec:post_attack} solves this difficulty, for now we only focus on the exact version to see the maximum capability of the method. Figure~\ref{fig:post_boost_synthetic} shows the empirical risk on the test vs. the number of submissions to evaluators. Here we assumed PostBoost is enjoying the complete knowledge of the logging $k$-NN RS model, including the vector representations of users and items. There are a few takeaways: PostBoost can consistently disrupt the evaluation process by $\sim5\%$ whether the evaluation mechanism is Kaggle or Ladder. Even on the first query, when no feedback from the evaluator is reported, PostBoost can guess the test set's labels better than the chance level. This gap exactly shows the value of the sequential information hidden in the test set.

It should be noted, that in both experiments with WBoost and PostBoost, we selected evaluation parameters manually to make the difference of Kaggle and Ladder mechanisms clear. We also reported results for the special case of $1$-NN RS with $exploration = 0.1$. However, similar results can be obtained by a wide range of $k$ and $exploration$.

\subsection{Evaluation on ML-100k}
The ML-100k ratings to movies range from $1$ to $5$. To use the ratings as binary labels, we associated ratings larger than $2$ with positive labels. Then we downsampled the prespecified test sets to a balanced test set consisting of $1000$ samples, preserving the time order of samples. In order to use $k$-NN posterior boosting attack, we need user and item vector representations. For simplicity, we used side information available for users and items as their representations. Specifically, we represent users with a three-dimensional vector of normalized age, sex, and occupation and represent items with a zero-one vector of genres. 

\begin{figure}
    \centering
    \includegraphics[width=0.8\linewidth]{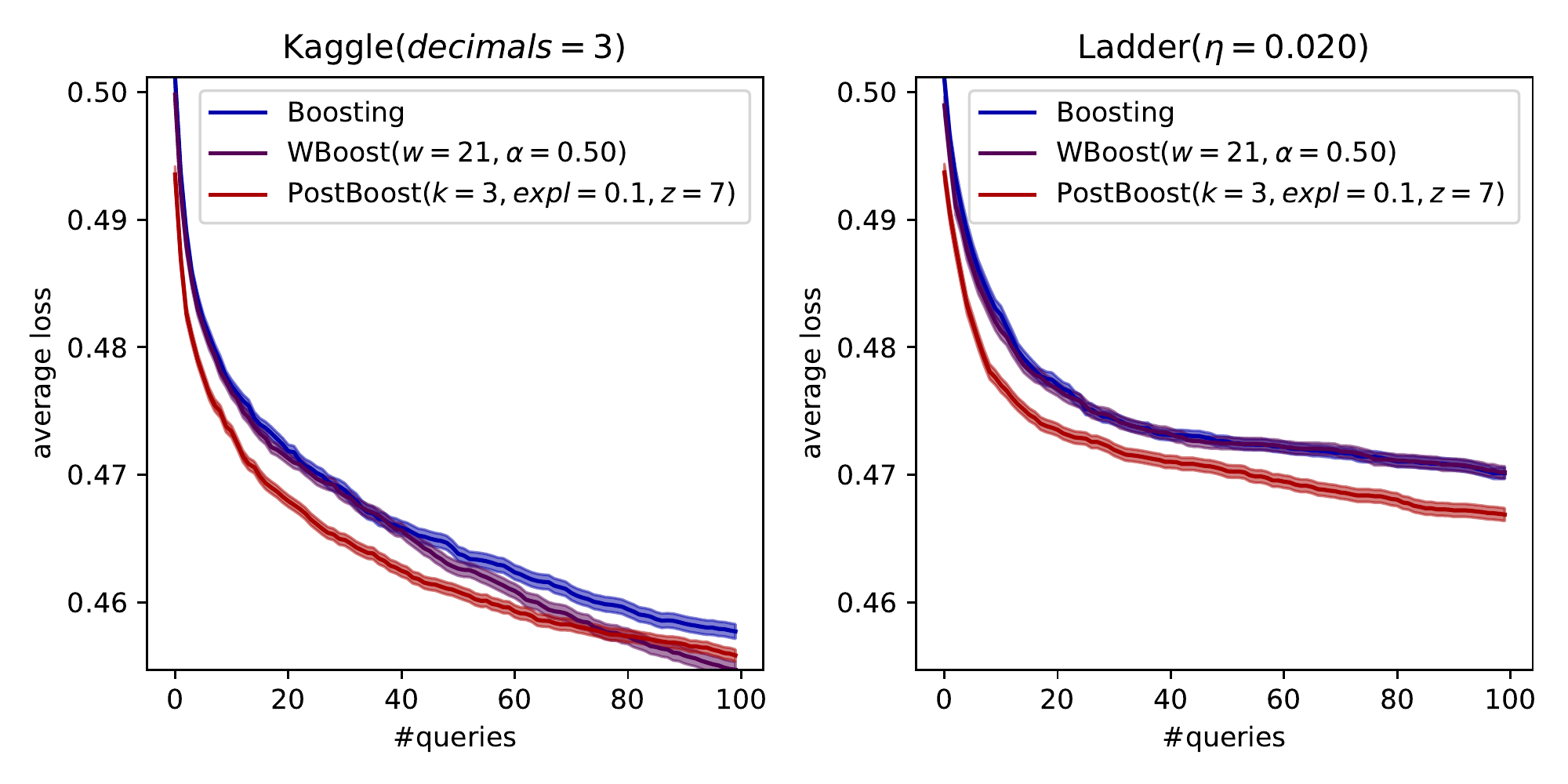}
    \caption{Empirical risk on the ML-100k's test sets while increasing the number of queries.}
    \label{fig:ml-100k}
\end{figure}

Figure~\ref{fig:ml-100k} shows the empirical risk on the test set while increasing the number of queries. To make the PostBoost computationally feasible, we used its approximate version here. Although we didn't know the actual algorithm behind the ML-100k collection, our $k$-NN PostBoost attacker can still exploit sequential information and disrupt the evaluation for relatively $\sim1\%$. Roughly speaking, this is a significant number as the performance of state-of-the-art methods on the ML-100k over the last $5$ years has only been improved for $\sim3\%$. So, disruption in the evaluation due to sequential information can change the leaderboard\footnote{https://paperswithcode.com/sota/collaborative-filtering-on-movielens-100k}. One can also see that WBoost is able to outperform the boosting attack by submitting more and more queries when the evaluation mechanism is Kaggle. This shows the distribution shift over time exists in real data and is informative about the test set. Finally, we should mention similar results hold for a wide range of $k$ and $exploration$, and we only chose current values for demonstration.

%% file: sections/discussion.tex
\section{Discussion}
In any dataset obtained through interactions, the knowledge of the purpose of that interaction (e.g., recommending favorable items) can inform us about the data we should expect. This extra knowledge can challenge the evaluation process. In this study, we focused on recommender systems as one of the most widely applied interactive systems and showed sequential information at the test time can be exploited to disrupt the evaluation process in both real and synthetic datasets.

This study has multiple implications. First, as our $k$-NN posterior boosting attacker suggests, samples of a sequence are not equally likely to be positive; some samples are harder to predict. This opens a question on the correct practice in evaluating two algorithms on sequential data. Second, although we focused on the evaluation process and test sets, the train sets also suffer from similar problems. Sequence-aware attackers can be used to obtain a joint distribution over training samples and this distribution can be utilized in unbiasing the training process. Finally, we used a simple $k$-NN RS to approximate the actual logging RS in real data to show that the sequential information matters. Future works can propose to learn the logging RS from the order of observations to extract more of this information.